\PassOptionsToPackage{unicode}{hyperref}
\PassOptionsToPackage{hyphens}{url}
\documentclass[
]{article}
\usepackage[margin=1in]{geometry}
\usepackage{needspace}
\usepackage{xcolor}
\usepackage{amsmath,amssymb}
\usepackage{iftex}
\ifPDFTeX
  \usepackage[T1]{fontenc}
  \usepackage[utf8]{inputenc}
  \usepackage{textcomp} 
\else 
  \usepackage{unicode-math} 
  \defaultfontfeatures{Scale=MatchLowercase}
  \defaultfontfeatures[\rmfamily]{Ligatures=TeX,Scale=1}
\fi
\usepackage{lmodern}
\ifPDFTeX\else
\fi
\IfFileExists{upquote.sty}{\usepackage{upquote}}{}
\IfFileExists{microtype.sty}{
  \usepackage[]{microtype}
  \UseMicrotypeSet[protrusion]{basicmath} 
}{}
\makeatletter
\@ifundefined{KOMAClassName}{
  \IfFileExists{parskip.sty}{%
    \usepackage{parskip}
  }{
    \setlength{\parindent}{0pt}
    \setlength{\parskip}{6pt plus 2pt minus 1pt}}
}{
  \KOMAoptions{parskip=half}}
\makeatother
\usepackage{longtable,booktabs,array}
\usepackage{caption}
\usepackage{calc} 
\usepackage{etoolbox}
\makeatletter
\patchcmd\longtable{\par}{\if@noskipsec\mbox{}\fi\par}{}{}
\makeatother
\IfFileExists{footnotehyper.sty}{\usepackage{footnotehyper}}{\usepackage{footnote}}
\makesavenoteenv{longtable}
\providecommand{\tightlist}{%
  \setlength{\itemsep}{0pt}\setlength{\parskip}{0pt}}
\usepackage{bookmark}
\IfFileExists{xurl.sty}{\usepackage{xurl}}{} 
\makeatletter
\@ifundefined{xmpquote}{}{}
\makeatother
\hypersetup{
  pdftitle={Evaluating Name-Only Directory Routing for One-Shot Code Search},
  pdfauthor={Manoj Bajaj},
  hidelinks,
  pdfcreator={LaTeX via pandoc}}

\title{Evaluating Name-Only Directory Routing for One-Shot Code Search}
\author{Manoj Bajaj}
\date{September 25, 2026}

\begin{document}
\maketitle

\begin{abstract}

Finding the right files is an early challenge for coding agents. We test
whether a language model can follow directory and file names to find
annotated code files missed by fixed lexical queries. Across 82 audited
issues from 11 repositories at pinned pre-fix commits, name-only
directory routing recovered 0.465 of gold files within eight candidates,
compared with 0.352 for FTS5 and 0.245 for a fixed full-issue
\texttt{rg} query. The paired gain over FTS5 was 0.113 (95\%
repository-cluster bootstrap interval, 0.053 to 0.168). Under a shared
16K-token context budget, routing delivered 0.443 of annotated lines
versus 0.246 for FTS5 on 55 cases with fully aligned annotations. At the
same eight-file limit, combining routing with FTS5 reached 0.491 file
recall, but its gain over routing alone was uncertain. An exploratory
flat path control reached 0.572 recall while using 24.6 model calls per
issue, compared with 8.9 for routing. Routing averaged 8.9 seconds per
issue; FTS5 took 7 milliseconds per query after a 0.9-second build. On
this cohort, directory routing added relevant file candidates to
one-shot lexical search, but the study cannot attribute the gain to
hierarchy or show that it improves issue resolution.
\end{abstract}

\section{Introduction}\label{introduction}

A coding agent can call \texttt{rg} repeatedly, but the first query is
often a behavioral description rather than an identifier. Search terms
may match documentation, changelogs, tests, and unrelated
implementation. The repository's directory structure supplies an
alternate route to candidate files: at each directory, a model selects
plausible children and may retain several branches. This paper asks
whether that route discovers \emph{additional annotated source evidence}
at a useful cost, not whether it makes text search obsolete. Our
contribution is a controlled, reproducible evaluation of when a
language-model-guided walk of repository names adds files omitted by
fixed lexical rankings, and whether those files survive an equal
context-selection budget. We do not introduce hierarchical code search.

The claim has two stages. First, a file-discovery method must nominate
gold files within a fixed candidate limit. Second, those candidates must
yield the required source lines within the coding agent's context
budget. A method can win the first stage and lose the second through
ordering, overly narrow windows, or noisy context. We therefore report
both stages rather than using span overlap or downstream answer scores
as a proxy for file discovery.

Our pilot motivates the question but does not answer it generally. On 24
previously inspected issues from Django and Svelte, name-only tree
routing found a larger fraction of gold files at eight candidates than
one fixed full-issue \texttt{rg} literal ranking (0.476 versus 0.219). A
stronger FTS5 control reached 0.295; tree and FTS5 both delivered
complete gold-line contexts for only 4/24 questions at 16K tokens. These
observations demand a new-repository held-out comparison with multiple
lexical controls and a shared final-context budget. We make no inference
from a selected three-file pilot example to general performance.

\Needspace{8\baselineskip}
\section{Research questions}\label{research-questions}

\begin{itemize}
\tightlist
\item
  \textbf{RQ1:} Does model-guided directory routing add gold
  files that tuned lexical search does not rank within the same top-k
  budget?
\item
  \textbf{RQ2:} Does adding tree candidates to lexical candidates
  improve complete file and line evidence under equal candidate and
  context budgets?
\end{itemize}

\section{Method and evaluation}\label{method-and-evaluation}

\subsection{Task and data}\label{task-and-data}

The unit is an issue statement paired with a repository at its pre-fix
base commit. The retriever receives the issue text and source tree only:
never annotated paths, gold line ranges, a solution patch, or post-fix
code. ContextBench~\cite{li2026contextbench} supplies
repository issues and annotated source regions. We pinned its revision
and Parquet checksum in a
\texttt{evals/fixtures/queries/contextbench\_paper.json}.
The 24 Django/Svelte pilot cases are excluded. The independent manifest
contains 87 issues at 87 commits from 11 other repositories in Python,
TypeScript, JavaScript, Go, C, and Rust. Repositories were selected by
eligible-case count within language strata; at most eight issues per
repository were selected by a hash ordering of instance ID. This tests
transfer across repositories but is not a random sample of software
projects. Source availability and common-corpus eligibility are audited
before scoring, with exclusions logged and never replaced. The exact
selection and audit rules are in the accompanying
\texttt{paper/study-amendment-001.md}. The source audit retained 82/87 cases. All five
exclusions had gold paths outside the common searchable-file policy,
including real source under directories named \texttt{build}. They
remain in the exclusion ledger. Among the retained cases, 55 had every
annotated span text align with the base-commit source after whitespace
normalization or exact-text containment. File discovery uses all 82;
line-level conclusions are shown both for all 82 and the 55-case
aligned-text sensitivity subset. The initial 171-case tree pass was
stopped at the author's request after the reduced subset had already
completed. Some interim tree outcomes were visible before the budget
amendment, but the first-eight-per-repository rule depends only on the
original hash-ordered manifest, not those outcomes. The study is not
externally preregistered or fully blinded.

\noindent\textit{Cohort by language, before and after the source audit.}
{\def\LTcaptype{none} 
\begin{longtable}[]{@{}lrr@{}}
\toprule\noalign{}
Language & Selected & Audited \\
\midrule\noalign{}
\endhead
\bottomrule\noalign{}
\endlastfoot
Python & 24 & 24 \\
TypeScript & 16 & 16 \\
JavaScript & 16 & 16 \\
Go & 16 & 13 \\
C & 7 & 7 \\
Rust & 8 & 6 \\
Total & 87 & 82 \\
\end{longtable}
}

\subsection{Retrieval arms}\label{retrieval-arms}

The tree arm walks the existing folder hierarchy. At each visited
directory, a categorical classifier receives the first 500 issue
characters and a menu of immediate child directory and file names. Each
directory option also includes up to 12 descendant names from at most
three levels below it. The classifier assigns probabilities and may
keep multiple plausible branches. The evaluated
implementation uses \texttt{typesafe/jev-1.13.0}, no generated directory
summaries, minimum and relative branch-probability thresholds of 0.01, a
32-call cap, and a 180-second per-issue time cap. It outputs at most 16
ranked file paths. This is a file-discovery method, not a complete
coding agent.

Here ``classifier'' means the TypeSafe System One \texttt{choice}
endpoint, not a classifier trained for this study. The request state is
\texttt{User question:} followed by the first 500 issue characters.
Each criterion is an immediate child path and name-only summary;
directory criteria additionally list up to 12 descendant names. A
\texttt{none\_of\_these} criterion is added. The instruction is:
``Choose the best route toward answer evidence. Directory options
summarize descendants, so select a directory when a descendant could
answer. Choose none only when the question is unrelated to every
option.'' The provider returns a probability for every criterion.
The JSON request has \texttt{questions.route.type=choice},
\texttt{instructions}, and \texttt{criteria=\{option\_i: text,\ldots,
none\_of\_these: text\}}; the adapter reads
\texttt{answers.route.probabilities} and rejects missing or out-of-range
values. We do not renormalize probabilities across menus. Menus are split at
254 options or 48,000 request characters; each option is limited to
800 characters. The \texttt{none\_of\_these} probability is not a
traversable child; the best-child threshold excludes it. Children with
probability at least the larger of
0.01 and 0.01 times the best child probability are retained. A stack
visits higher-scoring children first; tied children follow reverse
input order. Path rank is traversal order; branch probabilities are not
multiplied into a global file score. Single-child directories need no model call. The API
request supplies no temperature setting. HTTP 429/529 responses are
retried up to three times; other errors fail the issue. The archived
model identifier is \texttt{jev-1.13.0}; provider-side weights and
sampling behavior are not under our control.

\Needspace{8\baselineskip}
\begin{center}
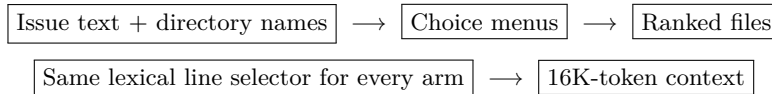
\small
\fbox{Issue text + directory names}\; $\longrightarrow$\;
\fbox{Choice menus}\; $\longrightarrow$\; \fbox{Ranked files}\\[5pt]
\fbox{Same lexical line selector for every arm}\; $\longrightarrow$\;
\fbox{16K-token context}
\captionof{figure}{Directory routing proposes files; the shared selector and
packer determine which source lines reach the context budget.}
\label{fig:pipeline}
\end{center}

Lexical arms search the same eligible UTF-8 files. The primary comparator
is in-memory SQLite FTS5 ranking from the first 500 issue characters,
chosen because it had the highest lexical recall@8 on the pilot. It
indexes paths and file content, searches an OR of unique non-stopword
tokens, and ranks with FTS5's BM25 function using path/content weights
2/1. An actual full-issue ripgrep replay searches case-insensitive
literal terms over the same files; its fixed file ranker weights term
rarity and path matches. It is one nonadaptive query, not a proxy for an
agent that can run successive searches or generate better keywords. The
hybrid applies reciprocal-rank fusion (rank constant 60) to tree and
FTS5 top-16 lists and truncates back to eight or sixteen files; it
receives no larger candidate allowance. We do not compare its one-shot
file ranking against Pi's multi-turn \texttt{grep}/\texttt{read} loop
using span recall; that would measure different systems and conflate
file discovery with section selection.

\subsection{Outcomes and statistics}\label{outcomes-and-statistics}

For issue q, let G be its annotated gold-file set and R(k) the first k
returned files. File recall is \textbar G intersect R(k)\textbar{} /
\textbar G\textbar; the main aggregate is the unweighted mean over
issues. We also count fully recovered gold-file sets, gold files found
by tree but absent from the same-size FTS5 prefix, and issues with at
least one such file. Since 16 of 82 audited issues have more than eight
gold files, complete recovery at eight candidates is impossible for
them; the ceiling is reported. File precision is secondary because the
intended use tolerates some extra files if required evidence is present.

Candidate files are not equivalent to delivered context. Every ranking
uses the same line selector: at most four lexical anchors per file, a
200-line radius around each, and merged overlapping windows. The windows
are packed under a 16K-token budget with the same tokenizer and order.
The selector tokenizes the full issue and every retrieved file, removes
stopwords and tokens of at most two characters, and scores each source
line by its count of distinct overlapping query terms. It takes the four
highest-scoring lines per file (earlier line wins ties), expands each
by 200 lines in both directions, and merges touching ranges. When a
file has no matching line, it offers the whole file. Ranges are packed
in file-rank order, with higher-scoring ranges first within a file;
an over-budget range is skipped and packing continues. The tokenizer
is \texttt{cl100k\_base}, and line numbers and paths count toward the
budget. The selector scans full retrieved files, but receives only the
issue text and those files. Gold paths, annotated line text/ranges,
patches, and post-fix code are never passed to retrieval or selection;
they enter only the subsequent scorer. The same selector is rerun for
every arm without arm-specific settings.
\Needspace{9\baselineskip}
\begin{verbatim}
for file in ranked_files:
  hits = top_four_matching_lines(issue_terms, full_file)
  ranges = merge(expand(hits, radius=200)) if hits else [full_file]
  for range in highest_hit_score_first(ranges):
    if used_tokens + numbered_block_tokens(range) <= 16000:
      include(range)
\end{verbatim}
We score annotated-line recall and complete line coverage after packing.
Because text annotations can differ from base-commit source, we show
line outcomes separately on the pre-retrieval aligned-text subset.
Paired differences use 10,000 repository-cluster bootstrap resamples
(seed zero): sample the 11 repositories with replacement and carry all
their paired issues into each sample. Per-repository results expose
concentrated gains. Setup,
query, model-call, token, cost, and wall-time measurements are reported
separately. Neither file nor line recall is issue-resolution success.
The reduced study does not repeat the tree run or measure the combined
hybrid pipeline wall time; it reports each measured arm's operations and
does not claim a fused end-to-end speedup. Prespecified descriptive
subgroups mark whether the issue explicitly mentions an annotated full
path or basename and whether its gold files span more than one
directory. These labels use the issue and gold metadata, never retrieval
outputs.

\section{Results}\label{results}

\subsection{Equal-budget retrieval}\label{equal-budget-retrieval}

Table~\ref{tab:retrieval} reports unweighted issue means. Every arm uses the same
candidate limit and context selector. The line-recall columns score
annotated lines actually included after the 16K-token pack on the 55
cases with aligned annotation text, not just lines in nominated files.
File columns use all 82 cases. Complete-file rates have ceilings of 66/82 at
eight candidates and 80/82 at sixteen because some issues annotate more
files than the limit.

\Needspace{11\baselineskip}
\captionof{table}{Equal-budget retrieval and delivered-line results.
File values use all 82 issues; L uses the 55 fully aligned-text issues.
All values are issue means. Full counts issues with every gold file in
the top-$k$. Fusion uses equal-$k$ RRF with rank constant 60.}
\label{tab:retrieval}
{\def\LTcaptype{none} 
\begin{longtable}[]{@{}
  >{\raggedright\arraybackslash}p{(\linewidth - 12\tabcolsep) * \real{0.25}}
  >{\raggedleft\arraybackslash}p{(\linewidth - 12\tabcolsep) * \real{0.125}}
  >{\raggedleft\arraybackslash}p{(\linewidth - 12\tabcolsep) * \real{0.125}}
  >{\raggedleft\arraybackslash}p{(\linewidth - 12\tabcolsep) * \real{0.125}}
  >{\raggedleft\arraybackslash}p{(\linewidth - 12\tabcolsep) * \real{0.125}}
  >{\raggedleft\arraybackslash}p{(\linewidth - 12\tabcolsep) * \real{0.125}}
  >{\raggedleft\arraybackslash}p{(\linewidth - 12\tabcolsep) * \real{0.125}}@{}}
\toprule\noalign{}
\begin{minipage}[b]{\linewidth}\raggedright
Arm
\end{minipage} & \begin{minipage}[b]{\linewidth}\raggedleft
File@8
\end{minipage} & \begin{minipage}[b]{\linewidth}\raggedleft
Full@8
\end{minipage} & \begin{minipage}[b]{\linewidth}\raggedleft
L55@8
\end{minipage} & \begin{minipage}[b]{\linewidth}\raggedleft
File@16
\end{minipage} & \begin{minipage}[b]{\linewidth}\raggedleft
Full@16
\end{minipage} & \begin{minipage}[b]{\linewidth}\raggedleft
L55@16
\end{minipage} \\
\midrule\noalign{}
\endhead
\bottomrule\noalign{}
\endlastfoot
Fixed full-issue \texttt{rg} & 0.245 & 12/82 & 0.116 & 0.328 & 16/82 &
0.116 \\
FTS5, first 500 chars & 0.352 & 17/82 & 0.246 & 0.421 & 19/82 & 0.252 \\
Name-only tree & 0.465 & 25/82 & 0.443 & \textbf{0.591} & \textbf{32/82}
& 0.457 \\
Tree + FTS5 & \textbf{0.491} & 23/82 & \textbf{0.499} &
0.571 & 30/82 & \textbf{0.504} \\
\end{longtable}
}

At eight candidates, tree had higher file recall than FTS5 on 33 issues,
lower on 15, and equal on 34. Its 61 gold-file occurrences absent from
FTS5's top eight appeared in 40 issues; FTS5 had 34 gold-file
occurrences absent from tree's top eight. Against fixed \texttt{rg},
tree had 82 such occurrences in 49 issues and \texttt{rg} had 29 absent
from tree. A further descriptive overlap count found 57 tree hits across
37 issues absent from both lexical top-eight lists. These are
\emph{issue--file occurrences}, not distinct files
across the corpus. Tree alone recovered eight more complete gold-file
sets than FTS5 at top eight. The fusion's higher mean recall did not
translate to more complete sets than tree: it recovered 23 versus 25.
The primary macro average weights issues equally. A secondary micro
average weights each annotated gold-file occurrence equally: at eight
candidates, tree found 108/391 (0.276) and FTS5 found 81/391 (0.207);
fusion found 120/391 (0.307). At sixteen, tree found 150/391 (0.384)
and fusion 151/391 (0.386). Thus the small tree--fusion ordering at
sixteen changes with the aggregation unit, while tree exceeds FTS5
under both.

\Needspace{16\baselineskip}
Table~\ref{tab:paired} gives prespecified paired differences; intervals resample
the 11 repositories, retaining all issues within each sampled repository.
A positive value favors
the left arm. The intervals describe sampling uncertainty across this
convenience cohort, not a general population guarantee.
A post-study exact sign-flip sensitivity over the 11 repository-level
paired tree--FTS5 contrasts gave $p=0.0078$ at eight files under
exchangeable repository signs; it is not a test of generalization to
unsampled repositories.

\Needspace{10\baselineskip}
\captionof{table}{Paired issue-mean differences and 95\% repository-cluster
bootstrap intervals. File columns use 82 cases; Line@8 uses the 55
aligned-text cases. Positive values favor the left arm.}
\label{tab:paired}
{\def\LTcaptype{none} 
\begin{longtable}[]{@{}
  >{\raggedright\arraybackslash}p{(\linewidth - 6\tabcolsep) * \real{0.2000}}
  >{\raggedleft\arraybackslash}p{(\linewidth - 6\tabcolsep) * \real{0.2667}}
  >{\raggedleft\arraybackslash}p{(\linewidth - 6\tabcolsep) * \real{0.2667}}
  >{\raggedleft\arraybackslash}p{(\linewidth - 6\tabcolsep) * \real{0.2667}}@{}}
\toprule\noalign{}
\begin{minipage}[b]{\linewidth}\raggedright
Comparison
\end{minipage} & \begin{minipage}[b]{\linewidth}\raggedleft
File@8
\end{minipage} & \begin{minipage}[b]{\linewidth}\raggedleft
Line@8
\end{minipage} & \begin{minipage}[b]{\linewidth}\raggedleft
File@16
\end{minipage} \\
\midrule\noalign{}
\endhead
\bottomrule\noalign{}
\endlastfoot
Tree -- FTS5 & +0.113 {[}0.053, 0.168{]} & +0.196 {[}0.125, 0.275{]} &
+0.170 {[}0.085, 0.252{]} \\
Tree -- fixed \texttt{rg} & +0.220 {[}0.106, 0.342{]} & +0.326 {[}0.242,
0.415{]} & +0.263 {[}0.148, 0.394{]} \\
Fusion -- FTS5 & +0.139 {[}0.092, 0.188{]} & +0.252 {[}0.181, 0.340{]} &
+0.150 {[}0.089, 0.219{]} \\
Fusion -- tree & +0.026 {[}-0.013, 0.069{]} & +0.056 {[}0.002, 0.126{]}
& -0.020 {[}-0.056, 0.018{]} \\
\end{longtable}
}

Fusion therefore improves this FTS5 ranking, but is not a demonstrated
improvement over tree alone. At top eight it raises file recall in 20
issues, lowers it in 13, and leaves it unchanged in 49 relative to tree.
At top sixteen, fusion lowers mean candidate recall slightly. Rank
fusion can displace a tree-selected gold file at a fixed limit;
combining candidate lists is not free. The aligned-text line difference
for fusion versus tree is +0.056 {[}0.002, 0.126{]}; this pre-retrieval
aligned subset is smaller and easier on file recall, so it does not establish a
robust general gain. On all 82 issues the annotation-sensitive
difference is +0.043, interval -0.006 to +0.101.

\subsection{Delivered context and annotation
sensitivity}\label{delivered-context-and-annotation-sensitivity}

The selector packed, on average, 12.7K tokens for tree@8, 14.2K for
FTS5@8, 14.9K for \texttt{rg}@8, and 13.8K for fusion@8, all within the
same 16K cap. Greater tree line recall was not explained by a larger
context allowance. Complete annotated-line coverage was nevertheless
uncommon: 14/82 issues for tree, 8/82 for FTS5, 5/82 for \texttt{rg},
and 15/82 for fusion. File discovery remains only the first step toward
sufficient coding context.

For the 55 issues whose every annotated text span aligns to the pinned
pre-fix source, line recall@8 was 0.443 for tree, 0.246 for FTS5, 0.116
for \texttt{rg}, and 0.499 for fusion. The paired tree--FTS5 difference
was +0.196 {[}0.125, 0.275{]}, and fusion--tree was +0.056 {[}0.002,
0.126{]}. These are the most trustworthy exact-line comparisons. The
predeclared all-82 line analysis remains in the archived artifact, but
is annotation-sensitive: 940/991 spans
aligned with the pinned source and 51 spans across 27 issues did not.
A post-study mechanical audit found the annotated text elsewhere in the
same file for 20 mismatched spans and did not find it in that file for
31; these labels do not establish why the dataset and source differ.
The mismatch is fixed before retrieval and cannot vary by retrieval
arm. Mismatched-case counts range from zero in GitHub CLI,
Transformers, and PonyC to six of eight in Material UI; every
repository's count and case IDs are archived with the audit. We do not
infer an exact-line improvement on the 27 cases from
the numeric range overlap alone. The aligned subset is not a random
sample of difficulty: tree file recall@8 was 0.531 there versus 0.331
on the 27 mismatched cases. We use it for annotation validity, not as
a representative replacement for all 82 issues.

\Needspace{8\baselineskip}
\captionof{table}{Mechanical alignment audit of 991 annotated spans.
Fifty-five issues align completely; 27 have at least one mismatch.
The mismatch categories are text-location checks, not cause labels.}
\label{tab:annotation-audit}
\begin{center}
\begin{tabular}{lr}
\toprule
Span result & Count \\
\midrule
Aligned at annotated range & 940 \\
Annotated text elsewhere in same file & 20 \\
Annotated text not found in same file & 31 \\
\bottomrule
\end{tabular}
\end{center}

\subsection{Repository variation and operational
cost}\label{repository-variation-and-operational-cost}

Tree exceeded FTS5 file recall@8 in nine of the 11 repositories, tied in
VS Code (0.13 each), and trailed in Serverless (0.16 versus 0.23). The
advantage was small for multi-directory gold sets (47 issues, +0.027
tree--FTS5) compared with single-directory sets (35 issues, +0.229).
This subgroup pattern is descriptive and may reflect both repository mix
and the difficulty of preserving several branches. It does not isolate
hierarchy as the cause of the gain. Nine issues explicitly name a gold
path; tree--FTS5 recall gain is +0.078 there versus +0.117 in the other
73. These small path-hint subgroups also do not establish a mechanism.
Table~\ref{tab:repositories} shows the repository-level
means, making both improvements and exceptions visible.

\Needspace{10\baselineskip}
\captionof{table}{File recall at eight candidates by repository.}
\label{tab:repositories}
{\def\LTcaptype{none} 
\begin{longtable}[]{@{}lrrrr@{}}
\toprule\noalign{}
Repository & n & Tree recall@8 & FTS5 recall@8 & Fusion recall@8 \\
\midrule\noalign{}
\endhead
\bottomrule\noalign{}
\endlastfoot
transformers & 8 & 0.78 & 0.59 & 0.78 \\
SymPy & 8 & 0.69 & 0.65 & 0.73 \\
Ansible & 8 & 0.60 & 0.45 & 0.57 \\
Material UI & 8 & 0.61 & 0.34 & 0.55 \\
VS Code & 8 & 0.13 & 0.13 & 0.13 \\
Serverless & 8 & 0.16 & 0.23 & 0.28 \\
NodeBB & 8 & 0.18 & 0.07 & 0.14 \\
GitHub CLI & 8 & 0.66 & 0.47 & 0.63 \\
Flipt & 5 & 0.29 & 0.23 & 0.42 \\
PonyC & 7 & 0.42 & 0.31 & 0.55 \\
Clap & 6 & 0.54 & 0.38 & 0.66 \\
\end{longtable}
}

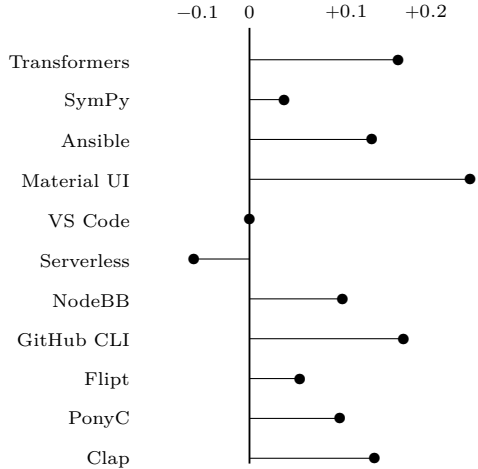
\begin{figure}[t]
\centering\scriptsize
\setlength{\unitlength}{1pt}
\begin{picture}(325,184)
\linethickness{0.7pt}\put(180,8){\line(0,1){164}}\thinlines
\put(148,174){\makebox(25,8){$-0.1$}}
\put(172,174){\makebox(16,8){$0$}}
\put(204,174){\makebox(25,8){$+0.1$}}
\put(234,174){\makebox(25,8){$+0.2$}}
\put(0,155){\makebox(135,9)[r]{Transformers}} \put(180,160){\line(1,0){56}} \put(236,160){\circle*{4}}
\put(0,140){\makebox(135,9)[r]{SymPy}} \put(180,145){\line(1,0){13}} \put(193,145){\circle*{4}}
\put(0,125){\makebox(135,9)[r]{Ansible}} \put(180,130){\line(1,0){46}} \put(226,130){\circle*{4}}
\put(0,110){\makebox(135,9)[r]{Material UI}} \put(180,115){\line(1,0){83}} \put(263,115){\circle*{4}}
\put(0,95){\makebox(135,9)[r]{VS Code}} \put(180,100){\circle*{4}}
\put(0,80){\makebox(135,9)[r]{Serverless}} \put(159,85){\line(1,0){21}} \put(159,85){\circle*{4}}
\put(0,65){\makebox(135,9)[r]{NodeBB}} \put(180,70){\line(1,0){35}} \put(215,70){\circle*{4}}
\put(0,50){\makebox(135,9)[r]{GitHub CLI}} \put(180,55){\line(1,0){58}} \put(238,55){\circle*{4}}
\put(0,35){\makebox(135,9)[r]{Flipt}} \put(180,40){\line(1,0){19}} \put(199,40){\circle*{4}}
\put(0,20){\makebox(135,9)[r]{PonyC}} \put(180,25){\line(1,0){34}} \put(214,25){\circle*{4}}
\put(0,5){\makebox(135,9)[r]{Clap}} \put(180,10){\line(1,0){47}} \put(227,10){\circle*{4}}
\end{picture}
\caption{Repository mean difference in file recall@8 (tree minus FTS5).
Dots right of zero favor tree; each repository contributes 5--8 issues.}
\label{fig:repo-differences}
\end{figure}

The tree made 729 Choice-model calls over 82 issues (8.89 per issue),
averaging 12.2K input and 1.93K output tokens, 8.93 seconds wall time,
and an estimated \$0.000512 per issue.
Its median/p95 routing times were 9.08/13.52 seconds. No tree issue
failed or hit the call or time cap.

\Needspace{6\baselineskip}
The estimate multiplies reported
input tokens by TypeSafe's Jev 1.13 rate of \$0.042 per million
\cite{typesafe2026models}, accessed September 24, 2026; output tokens
are free. It is not an observed bill. FTS5
index construction averaged 894 ms per pinned snapshot and its
first-500-character query averaged 7.5 ms; \texttt{rg} file enumeration
averaged 768 ms and full-issue search 1.68 seconds. FTS5 query
median/p95 were 4.8/20.6 ms and
\texttt{rg} search median/p95 were 0.71/6.10 seconds. These are observed
operation times, not comparable prebuilt-index versus cold-start
end-to-end latency claims. FTS5 and \texttt{rg} incurred no model API
charge.

\Needspace{6\baselineskip}
Fusion's end-to-end wall time was not measured; running both
sources and packing context cannot be represented by the FTS5 query time
alone.

\Needspace{18\baselineskip}
\subsection{Post-review path and query controls}
\label{post-review-controls}

After reviewing the original analysis, we added four controls on the
same 82 cases. These results are exploratory and did not change the
prespecified tree--FTS5 comparison. Path-only FTS5 indexes file names
without contents; a basename heuristic counts issue-token overlap in
path names. The flat Jev tournament shows every eligible path, in
bounded menus, to the same Choice model used by tree routing. Each
menu advances its top 16 paths until one ranking remains. It does not
use directory menus, but its instruction differs from the tree
instruction and it uses more calls. A two-search \texttt{rg} script
forms a second query from rare terms in the first search's matched
files, then fuses the two lists. It is fixed feedback, not an adaptive
coding agent. All four controls use the same file and context budgets
and the unchanged line selector. Table~\ref{tab:poststudy} compares
observed operating points: the flat tournament has a different
instruction and roughly 2.8 times as many model calls as tree. It is
not a compute-matched causal test of hierarchy.

\Needspace{12\baselineskip}
\captionof{table}{Primary arms and exploratory post-study controls. File
recall is the issue mean on all 82 cases; L55@8 is delivered-line recall
on the 55 cases with aligned annotation text.}
\label{tab:poststudy}
\begin{center}\small
\begin{tabular}{lrrr}
\toprule
Arm & File@8 & File@16 & L55@8 \\
\midrule
\multicolumn{4}{l}{\textit{Prespecified primary arms}} \\
Fixed full-issue \texttt{rg} & 0.245 & 0.328 & 0.116 \\
Content + path FTS5 & 0.352 & 0.421 & 0.246 \\
Name-only tree & 0.465 & 0.591 & 0.443 \\
Tree + FTS5 & 0.491 & 0.571 & 0.499 \\
\addlinespace[3pt]
\multicolumn{4}{l}{\textit{Exploratory post-study controls}} \\
Two-search \texttt{rg} & 0.260 & 0.316 & 0.209 \\
Path-only FTS5 & 0.191 & 0.237 & 0.214 \\
Basename overlap & 0.173 & 0.219 & 0.226 \\
Flat Jev tournament & \textbf{0.572} & \textbf{0.630} & \textbf{0.511} \\
\bottomrule
\end{tabular}
\end{center}

\Needspace{15\baselineskip}
\begin{center}
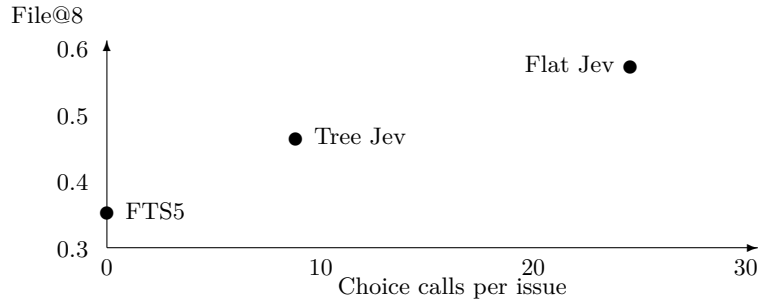
\small
\setlength{\unitlength}{1pt}
\begin{picture}(330,118)
\put(40,20){\vector(1,0){245}}
\put(40,20){\vector(0,1){78}}
\put(4,103){\makebox(52,9)[l]{File@8}}
\put(100,0){\makebox(140,10){Choice calls per issue}}
\put(34,8){\makebox(12,9){$0$}}
\put(113,8){\makebox(15,9){$10$}}
\put(193,8){\makebox(15,9){$20$}}
\put(273,8){\makebox(15,9){$30$}}
\put(5,15){\makebox(28,9)[r]{0.3}}
\put(5,40){\makebox(28,9)[r]{0.4}}
\put(5,65){\makebox(28,9)[r]{0.5}}
\put(5,90){\makebox(28,9)[r]{0.6}}
\put(40,33){\circle*{5}}\put(47,29){\makebox(50,10)[l]{FTS5}}
\put(111,61){\circle*{5}}\put(118,57){\makebox(62,10)[l]{Tree Jev}}
\put(237,88){\circle*{5}}\put(163,84){\makebox(68,10)[r]{Flat Jev}}
\end{picture}
\captionof{figure}{Observed model-call and file-recall operating points
on 82 cases. FTS5 uses zero model calls but has separate index and query
costs; tree and flat Jev differ in prompt, menus, and call counts.}
\label{fig:cost-quality}
\end{center}

Flat Jev exceeded tree at eight files by +0.107 issue-mean recall
{[}0.017, 0.224{]} in a post-study repository-cluster bootstrap;
at sixteen the difference was +0.039 {[}-0.038, 0.127{]}.
The flat control also delivered more aligned annotated lines at eight
files (+0.069 {[}0.006, 0.130{]} versus tree), though complete line
coverage remained uncommon. Its mean
24.6 calls, 156K input tokens, 30.6 seconds, and estimated \$0.00657
per issue compare with tree's 8.9 calls, 12.2K input tokens, 8.9
seconds, and estimated \$0.000512. These are observed operations and
estimated charges for the completed runs, not a call-matched test of
hierarchical traversal. A first timed-out attempt on one large flat
case made unarchived partial calls; that attempt is excluded from these
per-case means and disclosed in the amendment. The successful run had
one retried request and no scored failures.

The path-only lexical controls underperformed content-and-path FTS5 at
eight files. The two-search \texttt{rg} script improved fixed \texttt{rg}
file recall@8 by only +0.015 {[}-0.061, 0.083{]} and lost at sixteen;
its mean measured wall time was 2.67 seconds. Neither result implies a
trained model or an adaptive agent would behave the same way. The
flat result rules out a claim that the hierarchy itself is shown to
cause the original gain over lexical search. It leaves an operational
trade-off: tree routing found fewer gold files than flat Jev while
using fewer calls and tokens.

\Needspace{10\baselineskip}
\subsection{Illustrative gains and
failures}\label{illustrative-gains-and-failures}

In a GitHub CLI issue asking to hide the current branch in
\texttt{pr\ status\ -\/-repo}, the annotated file was
\texttt{pkg/cmd/pr/status/status.go}. Tree routing followed the command
hierarchy and ranked it eighth; neither FTS5 nor fixed \texttt{rg}
included it in the first eight. Conversely, for a VS Code issue about
command links rendered from Markdown, FTS5 ranked the annotated
\texttt{src/vs/base/browser/markdownRenderer.ts} fifth, while tree
routing followed workbench command files and missed it. Equal-k fusion
also missed the renderer in its first eight. These examples were
selected after aggregate analysis to show both a lexical-vocabulary
failure and a routing/fusion failure; they are not independent evidence
of average performance.

\section{Related work and boundaries}\label{related-work-and-boundaries}

Hierarchy-guided retrieval is not new:
PageIndex~\cite{pageindex} applies tree
reasoning to documents. Code-localization work exploits structure in
other ways. LocAgent~\cite{chen2025locagent}
uses a graph of code entities and dependencies for multi-hop
localization; repository-aware
file-path retrieval~\cite{yanuganti2025paths} trains a model to predict paths from
natural-language queries. These methods test different trade-offs from a
folder walk. Blink~\cite{blink} already
navigates codebase folders with Jev, so neither hierarchical code search
nor Jev use is claimed as a first here.

Lexical search is a serious baseline, not a straw man.
Better Call Grep~\cite{wang2026grep} studies
grep-like retrieval for repository-level completion. Our narrower
question is whether tree routing adds gold files to a fixed-budget
lexical candidate list when issue wording and source vocabulary differ,
and whether that addition survives context selection. A same-agent
comparison is needed before claiming an adaptive coding workflow
improves; offline list fusion alone cannot establish that. Nor do these
fixed-query baselines exhaust the space of intent-aware lexical search
or query reformulation.

\section{Limitations and conclusion}\label{limitations-and-conclusion}

On this cohort, name-only directory routing is a complementary source
of file candidates for one-shot, issue-driven code retrieval. At equal
candidate and context budgets, it found more annotated files and
delivered lines than the selected lexical baselines while using much
less model computation than the higher-recall exploratory flat path
tournament. The procedures
differ in prompt, menus, and calls, so the study does not isolate
hierarchy as the cause of the quality difference. Equal-budget
tree+FTS5 fusion improves over FTS5, but does not reliably improve over
tree and can displace tree-selected gold files from a fixed shortlist.

The primary study tests an issue's first fixed query, not an adaptive agent. A
multi-turn agent can reformulate searches, inspect stack traces, and
follow references. Therefore a tree-only file-recall advantage over one
fixed \texttt{rg} ranking would establish complementary candidate
generation, not superiority over grep as a tool or over Pi as an agent.
Fusion improves an agent only if the added files survive its context
selection and are actually used; we measure delivered lines but not
issue resolution.

The corpus policy excludes hidden files and directories commonly used
for generated outputs. Five selected cases lost gold files to that rule,
including real source under directories named \texttt{build}; those
exclusions are reported, not silently swapped. Folder names can also be
misleading, and the classifier may prune an essential branch. The
pre-fix gold text alignment check found 27 audited cases with at least
one mismatch; line-level results on those cases are less secure than
file-level results. The 11-repository convenience cohort and one
provider/model may not generalize to other repository layouts or query
styles. The original larger tree run was shortened after interim
outcomes were visible, although the reduced-case rule was independent of
those outcomes; the resulting evidence is weaker than a fully blinded
preregistered replication. The post-study flat Jev tournament used a
different instruction and more calls than tree. A call- and
prompt-matched comparison with repeated stochastic runs would be needed
to estimate the hierarchy's effect independently of compute allocation.
A chat-model gate would test the role of the Choice endpoint. The
scripted two-search \texttt{rg} control does not cover agent-led query
revision; code-symbol indexes and downstream coding agents remain
untested. Provider prices and latency can change, so the paper
reports observed calls/tokens and setup costs alongside dollar
estimates.

Directory routing thus provides additional annotated file candidates
that the selected one-shot lexical rankings omit. Its 8.93-second mean
latency and model charge make a blanket replacement for FTS5
unattractive; future systems should test when to invoke it, how to
preserve complementary candidates, and whether an agent actually uses
the extra evidence. This study establishes a retrieval result on a
fixed benchmark, not a measured improvement in software repair.

\Needspace{30\baselineskip}
\section{Reproducibility and data
availability}\label{reproducibility-and-data-availability}

The public code and data are pinned at
\url{https://github.com/manojbajaj95/fastindex/tree/5759fc9f0d4c84ec33c5e14b4366e3680194825e}.
That immutable commit records the case selection, source policy,
parameters, budget amendments, and reproduction commands under
\texttt{paper/} and \texttt{evals/}. It includes the pinned case
manifest, a ledger of all 82 retained case IDs, commits, query hashes,
gold ranges, and five exclusions, plus copies of the non-secret tree,
FTS5, \texttt{rg}, flat Jev, path-only and feedback controls,
gold-text audit, and original and post-study scored-result artifacts
under \texttt{paper/data/}.
Their SHA-256 checksums are in \texttt{paper/data/README.md}.
The audited holdout and source snapshots can be regenerated from the
pinned public commits. No API credentials or repository snapshots are
part of the paper artifact. The archived rankings support exact
re-scoring without model calls; raw Choice responses were not retained,
so they cannot be reconstructed from the ranked paths. A fresh model
run may choose different paths. The study archive therefore supports
analysis reproduction, not deterministic model replay.
From a fresh checkout with \texttt{uv}, \texttt{git}, and \texttt{curl},
run \texttt{sh paper/reproduce.sh}. This checks the dataset checksum,
materializes pinned source snapshots, and re-scores the archived
rankings without API credentials. The script contains the exact stage
commands and writes local results under \texttt{evals/results/}.

\paragraph{AI assistance and responsibility.}
OpenAI Codex assisted with implementation of the evaluation code,
execution and analysis of the experiments, and preparation of this
manuscript. The author is responsible for the study design, artifact
checks, interpretation, and final text. The study artifacts do not
record a single Codex model/version identifier across those sessions.
Codex was a research and writing tool; the evaluated routing decisions
were produced by TypeSafe Jev 1.13.

\bibliographystyle{plain}
\bibliography{references}

\end{document}